# Immunological determinants of clinical outcomes in COVID-19: A quantitative perspective

Elizabeth Krieger,[2*] Nicole Vissichelli,[1*] Stefan Leichtle,[3] Markos Kashioris,[1] Roy Sabo,[4] Don Brophy, [1] Xiang-Yang Wang,[5] Pamela Kimbal, [6] Michael Neale,[7] Myrna G. Serrano,[1] Gregory A. Buck,[5] Catherine Roberts,[1] Rehan Qayyum,[1] Daniel Nixon,[1] Steven Grossman[1] and Amir A. Toor.[1]

Virginia Commonwealth University Health System, Richmond, VA [1].

*- Equal contribution.

1- Department of Internal Medicine, 2- Department of Pediatrics, 3- Division of Acute Care Surgical Services, Department of Surgery, 4- Department of Biostatistics, 5- Department of Microbiology and Immunology, 6- Department of Pathology, 7- Department of Psychiatry and Statistical Genomics, Virginia Commonwealth University, Richmond, VA.

Correspondence, Amir A. Toor, Professor of Medicine, VCU, Massey Cancer Center, Richmond, VA. E-mail: amir.toor@vcuhealth.org

**Summary**

Severe acute respiratory syndrome coronavirus 2 (SARS-CoV-2) has a variable clinical presentation that ranges from asymptomatic, to severe disease with cytokine storm. The mortality rates also differ across the globe, ranging from 0.5-13%. This variation is likely due to both pathogen and host factors. Host factors may include genetic differences in the immune response genes as well as variation in HLA and KIR allotypes. To better understand what impact these genetic variants in immune response genes may have in the differences observed in the immune response to SARS-CoV-2, a quantitative analysis of a dynamical systems model that considers both, the magnitude of viral growth, and the subsequent innate and adaptive response required to achieve control of infection is considered. Based on this broad quantitative framework it may be posited that the spectrum of symptomatic to severely symptomatic presentations of COVID19 represents the balance between innate and adaptive immune responses. In asymptomatic patients, prompt and adequate adaptive immune response quells infection, whereas in those with severe symptoms a slower inadequate adaptive response leads to a runaway cytokine cascade fueled by ongoing viral replication. Polymorphisms in the various components of the innate and adaptive immune response may cause altered immune response kinetics that would result in variable severity of illness. Understanding how this genetic variation may alter the response to SARS-CoV-2 infection is critical to develop successful treatment strategies.

## Introduction

The coronavirus disease 2019 (COVID-19) pandemic of 2020 has created a challenge to humanity like none other in recent times. With its global reach, infections with the severe acute respiratory syndrome coronavirus 2 (SARS-CoV-2) have exacted a tremendous toll from populations around the world. This single-stranded RNA virus gains access to the host intracellular milieu through interaction with the angiotensin-converting enzyme 2 (ACE-2) which is expressed on a variety of epithelial and endothelial cells.[1 2] Not only is this novel virus associated with higher mortality than other respiratory viruses, such as influenza, but it also demonstrates a broader variation in its clinical presentation.[3 4 5 6] At disease onset, the symptoms are generally mild and restricted to the respiratory tract. Still, later, secondary viremia may involve other organ systems expressing ACE-2, including the cardiovascular, nervous, and renal systems.[7] Several patients experience an exaggerated immune response in the form of cytokine release syndrome.[8 9] Despite this, a vast majority of patients with COVID-19 may be asymptomatic or only mildly symptomatic, with the remainder experiencing a range of clinical manifestations, from moderate-to-severe, life-threatening.[7] Mortality rates are high in the elderly and those with comorbidities. Still, the young and otherwise healthy may also develop severe disease.[10 11] There is wide variability in case-fatality rates in different populations around the United States (**Figure 1A**) and the world, with rates as high as 10-13% seen in Italy, England, and Spain, but as low as 0.5-2% in Iceland, Norway and Japan (**Figure 1 B**). This variation in the global death toll from this infection, despite testing and reporting biases, may suggest a genetic component in disease susceptibility. This variation may represent differences in the relative proportions of tested versus non-tested populations, socio-economic, comorbid conditions, and differential public health practices. It may be postulated that the disparity in disease severity could be at least in part due to genetic differences in the innate and adaptive immune responses that individuals may mount to SARS-CoV-2.[12]

*Immune Response to SARS-CoV-2, a general overview*

COVID19 symptomatology may be due to a combination of pathogen and host factors. Progression to the more severe forms of disease not only dependent on pathogen factors such as viral inoculum and virulence but, most likely, also on host factors that lead to variable immune response. Patients with adequate immune function likely contain the virus without over-reaction.[11] However, in those with severe manifestations of COVID19, a dysfunctional immune response is frequently observed.[11 13] The immune system has co-evolved with a multitude of historical pathogens, inherent with redundancies to counteract pathogens which have developed multiple mechanism of immune evasion. This has led to the development of genetic differences and polymorphisms which may explain the variability of the immune response in some patients within the host immune pathway (**Supplementary Table 1**). It is, therefore, critical to explore the intricacies of the pathways of the immune response to understand the differences in disease severity further.

The immune response to viral infections is carefully orchestrated between the innate and adaptive responses, and various steps are required to attain control. Failure or suboptimal function of any of the components could lead to an inadequate response. The *innate* response initially recognizes pathogen-associated molecular patterns (PAMP) associated with viruses through pattern recognition receptors (PRR) in the tissue-resident antigen-presenting cells (APC), such as macrophages and dendritic cells (**Figure 2**). Examples of PRRs include Toll-like receptor (TLR) -7 & -13, which recognize single-stranded RNA,[14] [15] and RNA-sensing retinoic inducible 1 gene (RIG-1). [16] Nod-like receptors (NLR) also help by sensing damage-associated molecular patterns.[17] Polymorphisms in the TLR and its downstream signaling molecules such as MYD88 may alter responses triggered PAMP recognition.[18] [19] [20] The identification of viral pathogens by tissue macrophages initiates a cascade of cytokine secretion. These cytokines inactivate viral replication through type I interferons (IFN-$\alpha$ & IFN-$\beta$), and help activate the next line of innate immune defense, the natural killer (NK) cells, with their repertoire of activating and inhibiting receptors. The NK cell receptors include killer immunoglobulin-like receptors (KIR) and the lectin-like, NKG2 family of receptors, among others. KIR receptors recognize the down-regulation of human leukocyte antigen (HLA) molecules on infected cells and mediate cytotoxicity. NKG2 receptors recognize non-classical HLA molecules, such as HLA-E.[21] [22]

The innate immune effectors trigger a two-pronged HLA dependent adaptive immune response. The HLA genes exhibit extreme allelic polymorphisms and present viral peptides on host HLA molecules to T cells to trigger an adaptive immune response. The T cell response involves both HLA class I and II molecules, which engage cytotoxic T cell (Tc) and helper T (Th) cell populations, respectively. The latter promotes both the cytotoxic T cell proliferation, as well as, B cell immunity and generation of SARS-CoV-2 specific antibodies. Polymorphisms in the HLA molecules result in differential antigen binding and presentation, leading to variability in immune responses. Antibody production is the task of the B cell arm of the adaptive immune system, which with their unique B cell receptors, engage antigenic epitopes, either solubilized or directly presented by macrophages and dendritic cells.[23] [24] In response, B cells proliferate and differentiate into memory B cells and plasma cells facilitated by interaction with Th cells, establishing long term humoral immune response.[25] This humoral immune response is crucial to providing control for any viral infection, including SARS-CoV-2. A subset of patients may not develop long-lasting antibodies and remains unclear if they are at risk for recurrent infection.[26]

*HLA & KIR, the master regulators of immunity*

Polymorphisms in the HLA and KIR haplotypes may be responsible for the significant worldwide variability in the immune response to COVID19. In other viruses, genetic variation among HLA alleles and KIR confer differing susceptibility. For example, HLA-B*46:01 has been linked to the development and increased severity of SARS-COV-1.[27] HLA-A*02:05 may prevent HIV seroconversion, and HLA-B*52:01-HLA-C*12:02 had a protective effect on the progression of HIV in Japanese patients.[28] A study using in silico modeling to predict viral peptide-MHC class I binding affinity across all known HLA -A, -B, and -C genotypes, has shown that HLA-B*46:01 has the fewest predicted binding peptides to SARS-CoV-2. This finding suggests that patients with HLA-B*46:01 may be at increased risk for infection with SARS-CoV-2.[29] Thus, global haplotypes variation across populations may partly explain the difference in illness severity and case-fatality rates. A similar variety may be observed in the KIR haplotype frequencies. As an example, decreased KIR2DL2 expression has been linked to increased susceptibility to SARS-CoV-1 [30] and KIR2DL3 homozygosity in association with its ligand HLA-C1, is associated with reduced disease progression of hepatitis C virus.[31] The KIR2DL2 gene is present in 46% of Europeans and 80% of Ethiopians. 82% and 72% of these populations respectively are, at least, heterozygous for its ligand, C1. Similar HLA alleles and HLA epitope binding site effects have been observed in Dengue fever, transplant outcomes.[32 33 34 3536]

*T cell and B cell repertoire diversity and symptom severity in COVID-19 patients*

In addition to these innate immune pathways, adaptive immunity with its T and B cell repertoire diversity is critical to mount an adequate immune response to viral infections. The mammalian T and B cell repertoire diversity and ability to recognize pathogen-derived antigens, either directly (B cell receptors) or upon presentation by HLA (T cell receptors), is derived from a unique process of T and B cell receptor gene rearrangement. Each of these genetic loci (for the immunoglobulin heavy and light chain and T cell receptor $\alpha$ & $\beta$ loci respectively) are comprised of variable joining and diverse gene segments. These segments are recombined and further modified through the addition of non-templated nucleotides to yield an enormous repertoire of T and B cell receptor-bearing clones capable of recognizing the pathogens associated antigens with high precision and fidelity. A decline in repertoire diversity with age may create dominant oligoclonal T and B cell populations and predispose the elderly to develop symptoms with potentially higher severity of illness.[37 38 39] Further, inadequate T and B cell clonal responses in immunocompromised patients may put patients at risk for insufficient viral clearance, and a prolonged, severe, clinical course. As noted above, patients with COVID-19 develop significant lymphopenia, which may be due to a pan-T cell-suppressive effect. [40 41 42] A reduction in circulating T cells may represent migration to the site of inflammation, hyperfunction, or direct virally mediated lethality, and it is known that T cells also express markers of exhaustion in the face of a severe viral infection. [33 43]

*Inflammatory cytokines during COVID-19*

Cytokines are key signaling molecules that orchestrate cellular and humoral immune responses to viral infections. The cascade of pro-inflammatory cytokines is triggered by the engagement of TLR on macrophages and by NK cells not finding their cognate autologous inhibitory ligands on infected cells. This first wave of pro-inflammatory cytokines includes TNF-$\alpha$IL-1, IL-2, IL-6, IL-15, and IFN-$\gamma$. These stimulate Th cells to produce downstream cytokines, such as the pro-inflammatory IL-12, and 17, and the anti-inflammatory IL-4 and 10. These cytokines provide the critical third signal to trigger both Th and Tc responses to clear the infected cells. However, a generalized systemic inflammatory response may also be observed in viral infections. While optimal cytokine secretion will trigger an appropriate immune response, excessive cytokine release, i.e., “cytokine storm,” may develop in some patients with COVID-19, resulting in fulminant disease with high mortality. SARS-CoV-2 may also infect monocytes and dendritic cells, altering the cytokine expression patterns and contributing to lymphopenia observed. [44] Importantly, this results in overproduction of inflammatory cytokines IL-6 and downstream release of MCP-1, VEGF, and IL-8, eventually culminating in a cytokine storm. Secondary hemophagocytic lymphohistiocytosis may ensue in these patients. Patients with severe disease have higher levels of IL-1, IL-6, IL-8, IL-10, and TNF-$\alpha$ compared with patients with milder disease.[45] [46] These high cytokine levels are associated with lymphopenia, with a deficit of both Th, Tc subsets, including naïve T cells and regulatory T cells.[47] [48] Elevated cytokine levels are also associated with a decrease in HLA-DR expression.[42] In parallel, the NK cell population is also depleted, which may be the result of viral replication.[49] While those with severe disease have significantly higher SARS-COV-2 RNA load and lymphopenia has directly correlated with viral load,[50] these cytokine differences between moderately and severely ill patients may be due to polymorphisms in the cytokines involved. A quantitative approach relating differences in cytokine levels and polymorphisms in the immune response pathways may help identify patients at risk of severe disease.

*Immune response as a feed-forward process with signaling & effector components*

The immune response to infection may be broadly classified into a signaling component and an effector component, with the balance between the two determining the eventual outcome (**Figure 3**). The signaling function is determined by the cytokine and chemokine secretion by cells of the target tissues and the innate immune system in response to the infection. The effector component, on the other hand, is characterized by a pathogen-specific T and B cell response. The ability to maintain control of the virus and successfully recover from infection, relies on this sequential feed-forward loop nature of the signaling and effector components, with the virus attempting to inhibit these processes simultaneously. Mathematical models have been proposed to understand the quantitative nature of these processes. These models generally study the growth of viruses as an exponential function of time.[51] Viruses enter their host cells, grow exponentially in these cells, and are released into the surrounding milieu and infect an equal number of cells, where this exponential growth and release are repeated. Thus, viral replication exhibits exponential growth occurring in an exponentially rising number of target host cells until a limit is reached by target cell exhaustion. In SARS-CoV-2, because of the widespread expression of ACE-2, a large tissue reservoir is at risk of infection, amplifying the viral burden manifold over time. The corresponding immune response to the viral proliferation in the host has multiple components, and each of these components involves the growth of a population of immune effectors such as dendritic cells, NK cells, T cells, and B cells. The growth of each of these components will have to reach a threshold in an optimal period (e.g., by time, t1, t2, t3 or t, t', t'', t''' and so on for each cell type) to ensure timely and complete clearance of the virus (**Figure 4**). As the first line of defense, interferons slow down viral replication. Concomitantly, NK cells and cytotoxic T cells proliferate and kill infected cells but at different rates than the virus, driven by their receptor affinities.[52] For NK cells, this represents the balance of inhibitory and activating signals from KIR and NKG2 family of receptors. In some individuals with a KIR B haplotype and a larger component of activating receptors, this may be a more effective process as opposed to those with a dominant component of inhibitory receptors. For the T cell responses, the ability of the HLA haplotype of the individuals to present SARS-CoV-2 derived peptides will be critical. This encompasses both the antigen-binding affinity of the HLA molecules in the host as well as antigen abundance. The requisite T cell response will be proportional to this antigen affinity and the affinity of the T cell receptor to the viral antigen-HLA complex. While the NK and T cell subset proliferation may catch up with an exponentially rising number of infected cells,[53] cellular immunity may eventually fail under the pressure of rapidly replicating SARS-CoV-2, with widespread tissue involvement due to extensive ACE-2 expression. Thus, in addition to an efficient cellular immune response, a robust humoral immune response is essential to control and eliminate the infection eventually. The humoral response is characterized by a proliferating B cell and plasma cell population. Each plasma cell makes large amounts of pathogen-specific antibodies, which finally allows the host to match the growth rate of the virus and neutralize the viral particles being generated. Failure or suboptimal rate of any one of these pathways will lead to inadequate immune response and delay in viral clearance. The suboptimal response will potentially lead to earlier components in the pathway over-compensating for the failure of downstream mechanisms, caught in a feedback loop leading to phenomenon such as cytokine storm. The immune pathways are all susceptible to genetic polymorphisms that have functional consequences, such as variability in cytokine expression, antigen-binding affinities, the strength of receptor ligation and downstream signaling. [54] [55] [56] Thus, functionally consequential polymorphisms in these interconnected immune pathways may impede the development of an optimal immune response to COVID-19.

*Estimating the risk of severe illness with COVID19*

The immune response to viral infections is a multi-step, precisely coordinated process, which results in viral clearance through initial innate and later adaptive immune mechanisms. This has been modeled mathematically to a high level of precision using ordinary differential equations.[57 58 59] The immune system behaves like a dynamical system when both T cells and NK cells are considered.[60 61 62 63] Small changes in parameter values, for instance, in SARS-CoV-2 antigen – HLA binding affinity or the initial IL-6 or IFN-? levels produced in response to infection, may have a profound impact on the eventual clinical outcome.[64] Thus, polymorphisms in critical immune response genes may alter the clinical outcome in a significant manner, mainly when they are considered together in a comprehensive mathematical description of the total immune response. Quantifying the number of polymorphisms across variables can, therefore, be used to assess differences in disease risk. As such, it may be assumed that there are a set of variables (innate and adaptive) that confer an advantage in terms of viral clearance without cytokine storm and reduce the risk of severe disease, as compared to the set of variables with no common elements (**Figure 5**). In patients who have common elements across the entire set of immune response determinants (overlapping sets), the risk of severe disease may be equivalent when adjusted for age and comorbid conditions such as obesity, lung disease, and diabetes. Among patients who either have very few or no overlapping elements, the risk of disease may be significantly different.

*Dynamical systems modeling of the immune response to viral infections*

These considerations become evident in a thought experiment to model the growth of any virus. If $r_v$ is the viral growth constant (number of virions replicated, for each cell infected, for each iteration of the growth process), then the total viral burden *V* at time *t* can be given by the equation

$$V_t \approx r_v H(IE)t$$

Where *H* is the target tissue reservoir, and *IE* is the infection efficiency for that virus (proportion of virions that end up infecting a new cell). With an initial inoculum of 1000 viral particles, 100 cells (*H*) get infected (*IE*=0.1 or 10%); in the second iteration 100000 virions will be produced with 10,000 cells infected, and it this keeps growing over time *t* until the limit of *H* is reached. Interferon (IFN) would directly suppress *V*, by a fraction, $s_c$ (cytokine-induced suppression), starting at a later time $t'$ ($t' < t$), and depending on how IFN much is produced, modifies $V_t$ to $V_{t(IR)}$ (diminished viral burden following the immune response)

$$V_{t(IR)} \approx r_v H(IE) t \frac{1}{s_c t'} \text{ ......... [1]}$$

Along with IFN production, a cascade of cytokine signaling will also be initiated, by the tissue-resident dendritic cells (DC) as well as the innate immune cells migrating into the tissues,

$$DC \approx (\phi_0 + \Delta\phi) e^{ct'} V \text{ ........[2]}$$

Where $c$ is the unit production of cytokine per unit increase in $V$. $\phi_0$ are the tissue-resident macrophages, and $\phi$ are the macrophages/inflammatory cells that migrate into the tissues. $DC$ then is the dendritic cell ($\phi$) production of cytokines in response to $V$. As $V$ gets larger, $DC$ gets higher with $t$ and constitutes the feed-forward loop depicted in **Figure 3** and **Figure 6**. Unless the overall feedback loop depicted in **Figure 3** slows down the growth of $V$ over time, this can lead to cytokine storm. It is also important to note that $DC$ is a *vector matrix* made up of many different cytokines, which all work on their respective receptors in the different target cell populations.

The cytokines will trigger an NK cell response, which will slow down $V$. Of note, this variable, and others, represent growth or change over time and is more accurately referred to by $dV/dt$. Still, for simplicity, we will simply consider absolute value $V_t$. NK cell growth and response may be considered as a function of the sum of activating and inhibitory receptors, the KIR and the NKG2 family of receptors. These are modeled as vector-operator equations, where the NK cell with its KIR molecules constitutes a 'vector,' and the target with its KIR-ligand molecules, an operator. The 'operator' modifies the 'vector' upon interacting with it; for example, it may either inhibit or activate it. Each individual KIR-KIRL interaction may be described as follows; if an inhibitory KIR (iKIR) on the NK cell encounters a ligand on its target, this results in an interaction which may be scored, $(-1) \times (1) = -1$, this will give the NK cell an inhibitory signal, assuming constitutively active basal state for NK cells; if there is no ligand for an inhibitory KIR, i.e., missing KIRL (mKIRL), the interaction score will be $(-1) \times (-1) = +1$ because of the abrogation of the inhibitory signal, and finally, activating KIR (aKIR) interacting with its ligands will be scored, $(1) \times (1) = +1$, when the ligand is present, and $(1) \times (0) = 0$, when the ligand is absent since no signal is given. Each of these different scores constitute a distinct component of the total KIR effect on individual NK cells expressing them, and while aKIR and iKIR function independently the cumulative effect calculated by taking their sum, determining the eventual outcome of NK cell- target interaction. Activating KIR are not required for a robust NK cell immune response in most instances, evident as individuals with haplotypes containing no functional aKIR are common and even people with haplotypes that include aKIR have NK cells in their NK cell repertoire without any aKIR present.[65] Depending on the degree to which the virus down regulates HLA class I molecules to evade Tc mediated killing, it may lead to iKIR-missing KIR ligand mediated killing or may not through viral immune escape mechanisms such as, no HLA class I down regulation or HLA decoy expression. Though viral evasion through decoy HLA expression can lead to NK cell medicated viral killing as in m157 protein expression by mice with mouse CMV infection. M157 has been shown to strongly associate with an activating mouse LY49 receptor (analogous to KIR) and confer host protection against MCMV.[66] So, the NK cell effect may be summarized by

$$NK_{t''} \approx \frac{(Ne^{\Sigma(\mathrm{KIR-KIRL})}) * NK_0}{(Ne^{\Sigma(\mathrm{KIR-KIRL})} - NK_{t''-1})(e^{-r(NK)t''}) + 1}$$

In this equation, $NK_t$ is the number of NK cells at time $t$, $N$ is the growth constant for NK cells, Σ KIR-KIRL is the sum of activating and inhibitory KIR-KIR L interactions described above. The growth exponent $r(NK)$ is the cytokine-driven and intrinsic growth rate of NK cells. $NK_t$ will reduce $V_t$ by a certain fraction as well, so equation 1 becomes,

$$V_{t(IFN+NK)} \approx r_v H(IE) t (\frac{1}{s_c t' + NK_t t''}) \ldots\ldots [3]$$

Here time, $t''$ is a later period in time at which the NK cell effect becomes manifest, it is later than the time, $t$ when viral proliferation begins. This means that the impact of an immune checkpoint triggered later than $t$ (i.e., $t'$ and $t''$) will require for those processes to outpace the viral proliferation, which already had a lead on these in time.

The resulting viral infection and cytokine secretion trigger the adaptive immune response. In the adaptive response, the antigen-presenting cells (APC) present viral antigens on HLA molecules to the Tc and Th cells. The unique T cell receptors (TCR) engage the cognate HLA-viral antigen complex. The binding affinity of these antigens to the HLA molecules varies over a broad spectrum, as does the affinity of the TCR for their cognate antigen-HLA complexes. A library of viral peptides may thus be presented on HLA molecules and trigger a polyclonal T cell response. A system of vector-operator matrices with the logistic equation of growth governs the behavior of each T cell clone.

$$Tc_{j_{t'''}} \approx \frac{(m_i Te^{a_i I * TCR_j}) * Tc_{j0}}{(m_i Te^{a_i I * TCR} - Tc_{jt'''-1})(e^{-r(T)t'''}) + 1}$$

This equation describes the growth of a single clone of CD8+ Tc cells, where $a_i I$ is the binding affinity of the $i^{th}$ SARS-CoV-2 antigen to HLA class I molecules, $m_i$ is the expression level of the protein from which the $i^{th}$ peptide is derived, and TCR is the binding affinity of the TCR of *jth* Tc cell clone to the antigen-HLA complex. The coefficient T represents the growth constant for T cells. Analogously, CD4+Th cells will have $a_p II$ as the afiinity of the $p^{th}$ antigen for specific HLA class II molecule, which along with the affinity of the $q^{th}$ TCR for the antigen-HLA complex, and the expression of this protein determine the final cell number for $q^{th}$ Th cell clone upon stimulation by target $p^{th}$ antigen. It is to be noted that the starting time here is different, $t''' < t$. Each individual depending on their HLA types will thus have a different library of SARS-CoV-2 peptides that they will present to their Th and Tc cells. Further, Th cells modulate Tc and B cell proliferation and differentiation, and therefore their role may be described as a vector transformation function, changing the $j^{th}$ T cell clone at time $t'''$ from $Tc_{j_{t'''}}$ to $Tc_j{}'_{t'''}$.

$$Th_{q_{t'''}} \approx \frac{(m_p Te^{a_p II * TCR_q}) * Tc_{q0}}{(m_p Te^{a_p II * TCR_q} - Tc_{qt'''-1})(e^{-r(T)t'''}) + 1}$$

$$Tc_j{}'_{t'''} \approx Th_{q_{t'''}} * Tc_{j_{t'''}}$$

As noted, these libraries of antigens bound to HLA molecules are also modeled as *operator matrices*, which tend to vary between individuals with different HLA types and the T cell clones are modeled as *vector matrices*. The T cell responses are amplified by the cytokines secreted by the APC growing and generally tend to peak before plateauing. While the Tc are capable of precisely recognizing infected cells and destroying them, here, the growth is occurring on a different scale of magnitude altogether, compared to the growth and infection rate of virions. This differential magnitude occurs because each of the T cell clones starts with a very low number of cells (theoretically a single cell) when it first encounters cognate antigen.

The adaptive immune component is equipped to deal with rapidly replicating viruses through B cells. Each B cell makes millions of virus-neutralizing antibody molecules, so once B cell immunity sets in, it may effectively compete with viral growth through antibody secretion. There, too, an element of time is crucially involved, as the B cells with the appropriate B cell receptor (BCR) encounter the viral epitopes and get selected for expansion. This is, once again, a polyclonal process, with the affinity of the $k^{th}$ antigen for the $l^{th}$ BCR driving the B cell clonal expansion, as in the case of T and NK cells. The expanded B cell clones under the influence of Th cells undergo somatic hypermutation to further fine-tune the BCR affinity for the $k^{th}$ antigenic epitope yielding a modified $l^{th}$ BCR'. As these B cells proliferate, they produce large amounts of antibodies against the antigenic epitope $k$, $I_l$, which over time, catches up and neutralizes $V$, even though antibody secretion had started at time $t''''$ much later than $t$, when viral proliferation had started.

$$V_{t(IR)} \approx r_v H(IE)t\left(\frac{1}{s_c t' + DC\ (NK_t t'' + (\sum_1^p Th_q\, t'''\left(\sum_1^i Tc_j\, t''' + \sum_1^k B_l I_l\, t''''\right)))}\right)$$

The above equations outline the two possible innate and adaptive immune response scenarios to a viral infection that unfolds as time progresses, from $t \rightarrow t' \rightarrow t'' \rightarrow t''' \rightarrow t''''$. If this response controls the viral infection such that the actual viral load, $V_t \approx V_{t(IR)}$, with the viral load optimally eliminated, the signaling cascade, $DC$ (**Equation 2**), will be dampened and clinical course is optimal. If, on the other hand, $V_t > V_{t(IR)}$, persistent $V_t$ will continue to drive the $DC$ in a feed-forward loop culminating in a cytokine storm. In light of the above discussion, another mechanism for cytokine storm in the former situation (with optimal, timely viral control) will be autologous tissue destruction perpetuating the cytokine cascade in a different feedback loop. It is important to recognize that DC is a matrix, where many different cytokines with both suppressive as well as growth promoting effects are represented.[67] Thus the overall effect of DC on T cell or B cell growth is not always in the positive direction, rather depending on the cytokines dominating in the milieu at any given time one may observe growth suppression, a phenomenon commonly seen in severely ill COVID19 patients. The whole immune cascade may thus be mathematically described as a system of matrix vector operator equations. This schematic depicted in **Figure 6**.

*Inferences from the dynamical systems modeling of viral disease*

This exercise demonstrates the multiplicity of responses involved in controlling infections in general, as it is in this particular case. Further, they indicate the many redundant pathways at work in the immune response to the same. It is also imperative to note that pathogens have evolved mechanisms to evade the immune response at multiple checkpoints. Because of genetic variation, not all individuals are equally well endowed with the ability to control infection. A multi-targeted approach may be needed to overcome the redundancies encountered. The notion of performing randomized trials with a single intervention tested at a time is at the heart of medicine, and agents are failing such trials are often discarded for not having met efficacy endpoints. The thought experiment above suggests that for a pathogen that has the capability of outpacing or disabling several different immune mechanisms, a combination of many agents and modalities may be required to control the disease. Attention to this aspect of infectious disease management needs to be given in future trial design.

A quantitative approach, such as outlined here, may be used to identify variation in immune response and genetic components that may be responsible for the range of illness severity that is observed in COVID-19. Obtaining a better understanding of the differences in the host response is vital to identify those at risk for developing severe disease and target treatment strategies.

**Figures.**

**Figure 1. A**. COVID-19 population-adjusted incidence in US states as of 4/7/2020. (*E. Krieger, personal communication, source:* https://covid19.healthdata.org/united-states-of-america). Adjusted for population and each state graph started when it reached 0.5 persons/100,000 people. (Y-axis: Total COVID19 cases/state (log scale), X-axis: Time in days). **B.** The case fatality rate of COVID19 across the world. Note Log-Log scaling, and different countries occupy different slopes corresponding to variable case-fatality (*source: https://coronavirus.jhu.edu/data/mortality)*.

**1A**

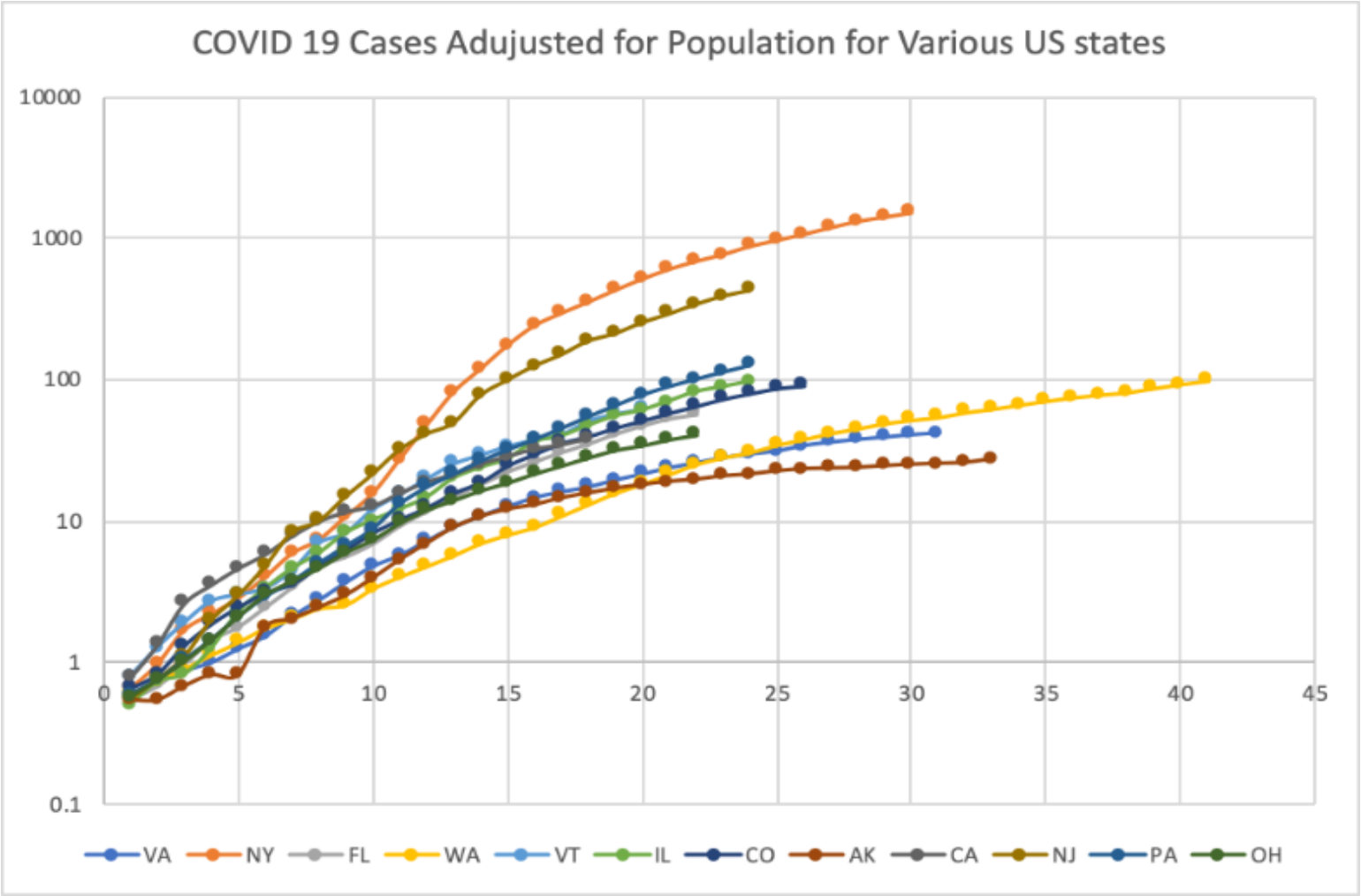

**1B**

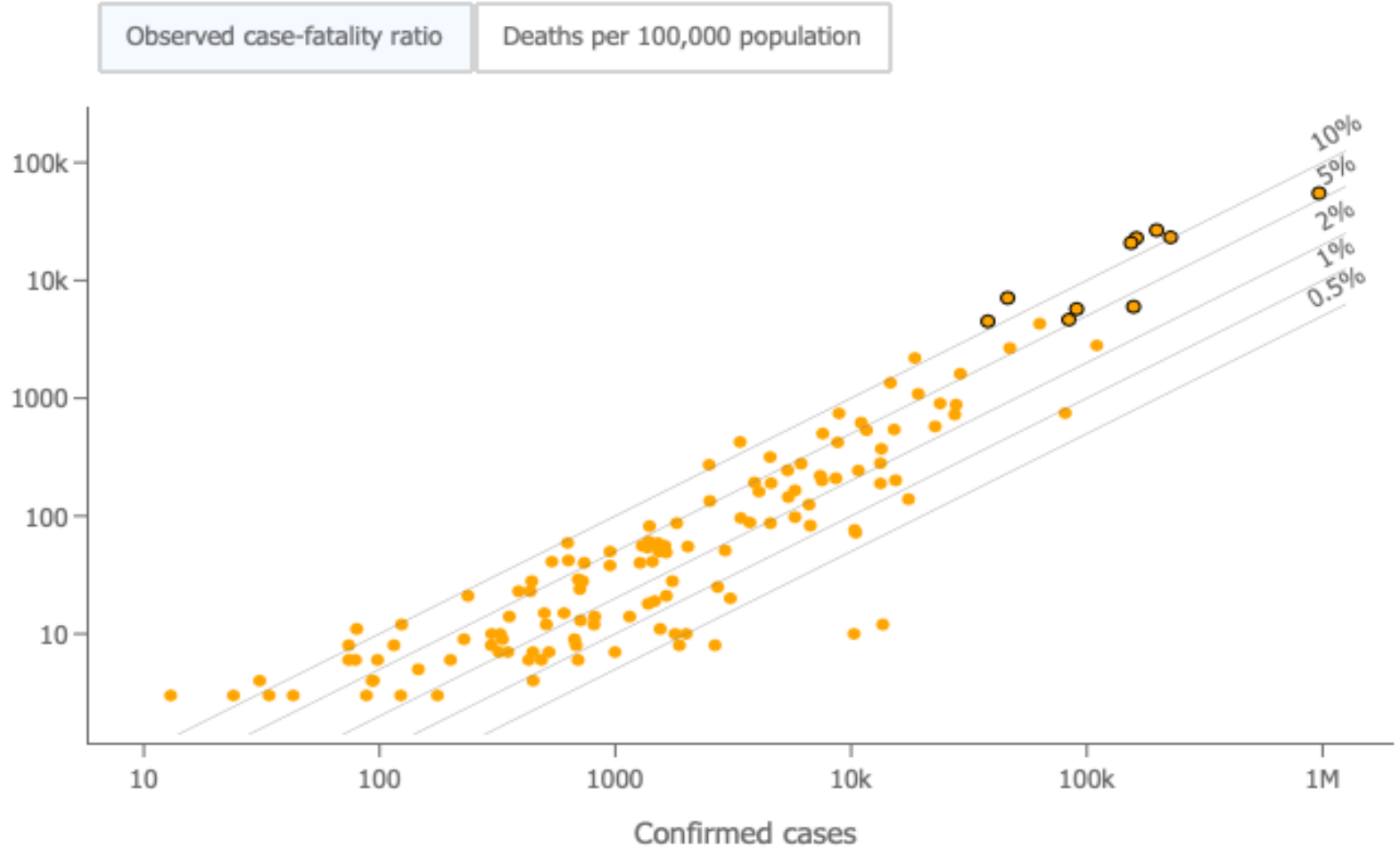

**Figure 2.** A schematic of the overall immune response to a viral infection in a healthy individual. Box (dashed line) represents the lymph node. The immune response destroys cells with disrupted margins. Time goes from $t_0$ to $t_5$ in this optimal scenario. APC: Antigen-Presenting Cell, DC: Dendritic Cell

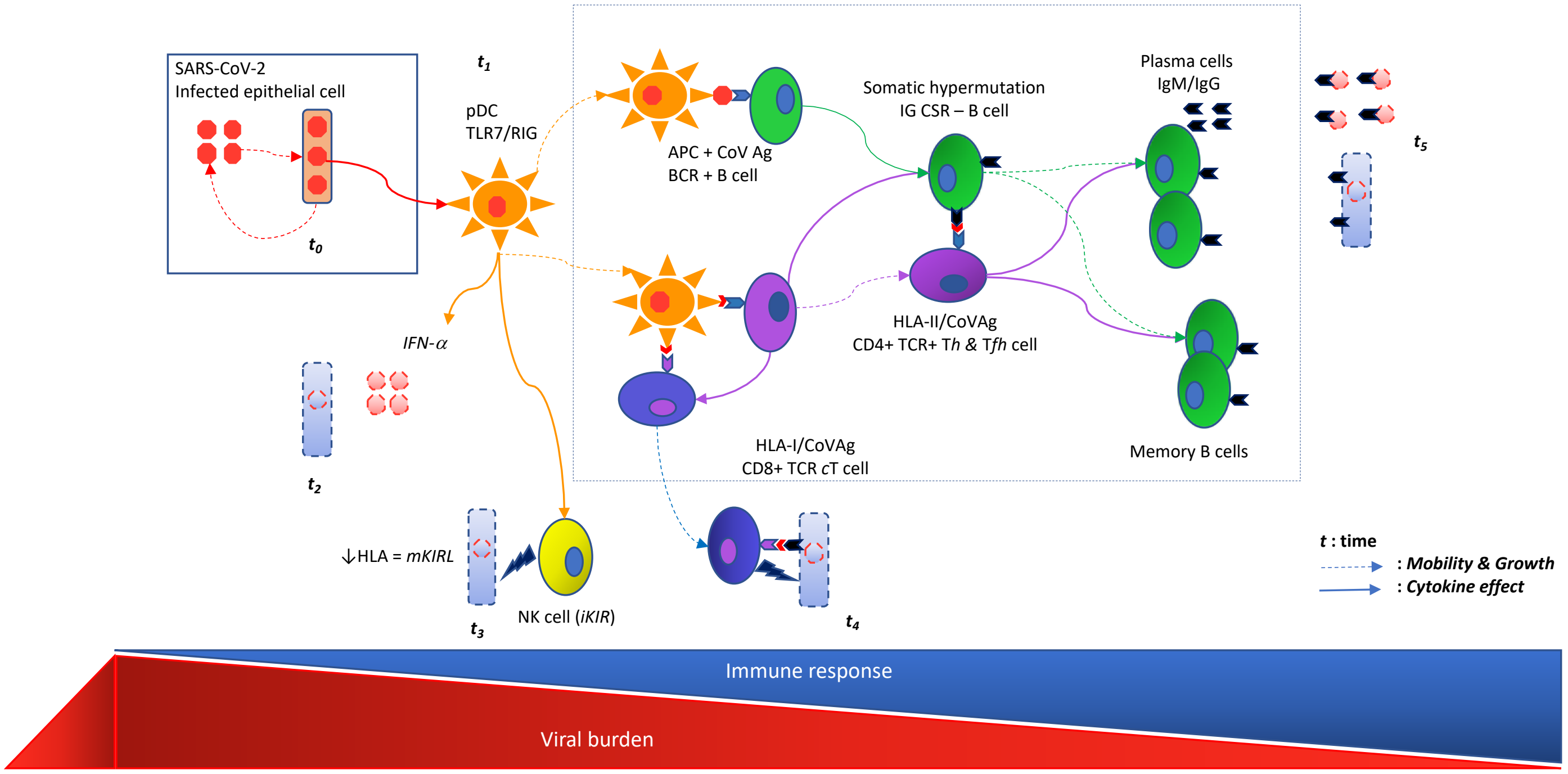

To clear SARS-CoV-2: Immune response > Viral burden over *t*

**Figure 3.** The magnitude of the immune response to viral burden over time and its components determine clinical outcome.

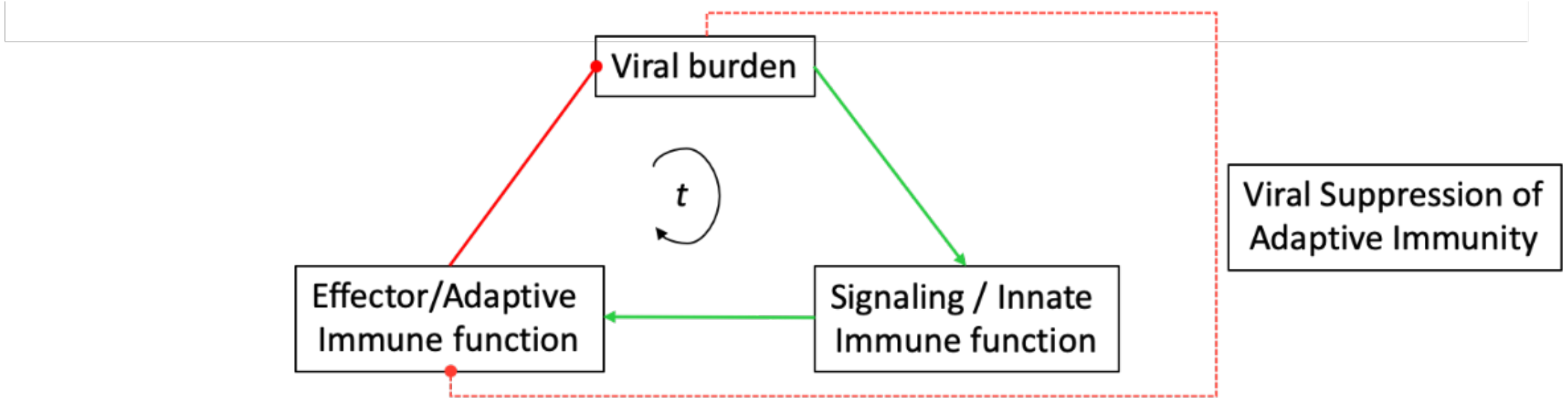

**If - Effector function > Viral burden over time *t*, Signaling function is downregulated**

*Effector function persistently > Viral burden → recovery from disease*

**If - Effector function < Viral burden over time *t*, Signaling function is upregulated**

*Effector function persistently < Viral burden → excessive cytokine release or overwhelming disease*

**Virus may downregulate adaptive immune response**

**Figure 4.** Quantitative model of antiviral immune response, demonstrating the evolution of innate and adaptive immune responses. $\sum_0^x Tc$ , $\sum_0^y Th$ & $\sum_0^{z'} B$ denote the sum of *x*, Tc; *y*, Th and *z'*, somatically hypermutated B cell clones.

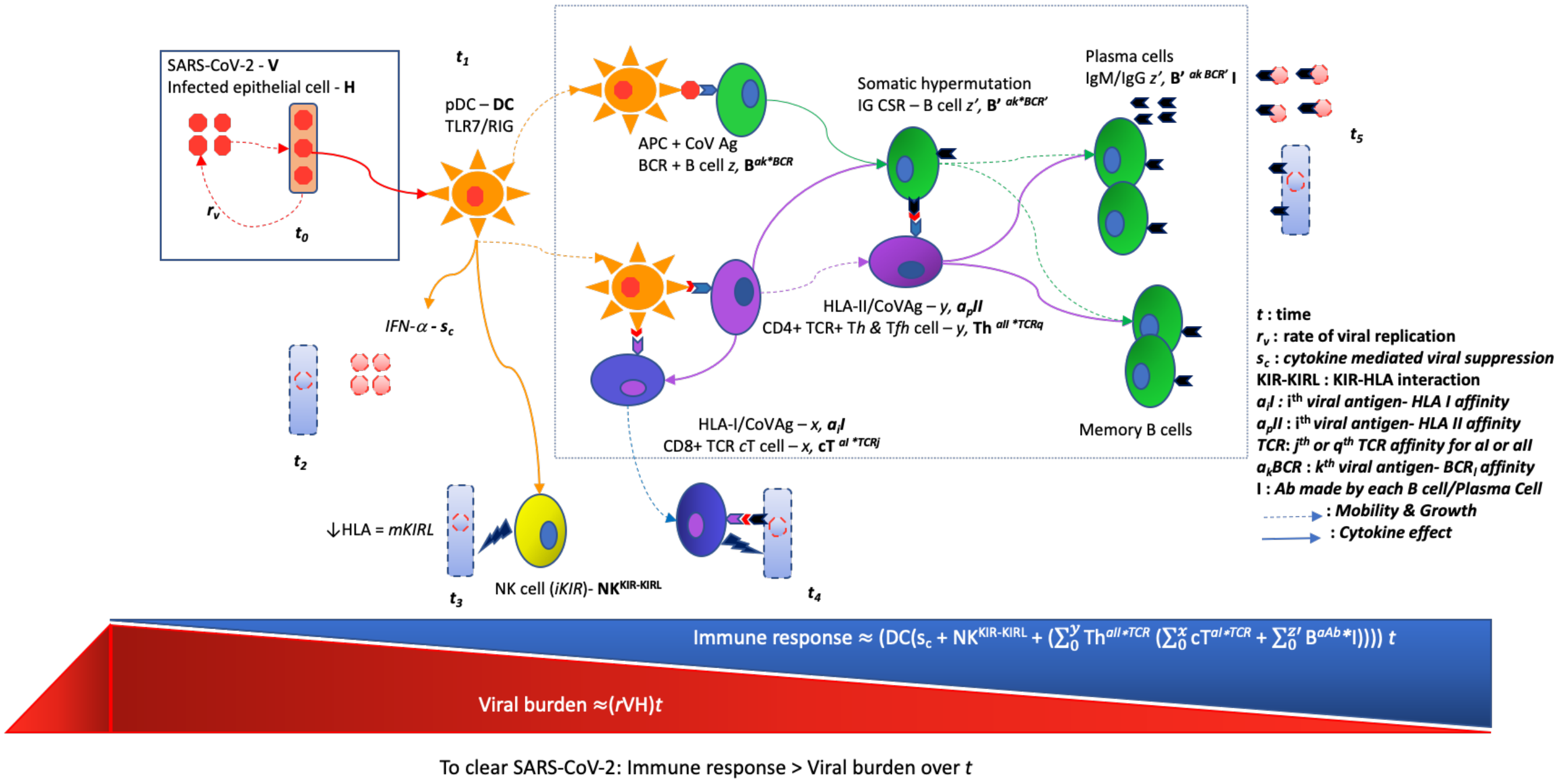

**Figure 6.** Vector operator model to illustrate the quantitative

**A.** Viral replication drives the signaling function

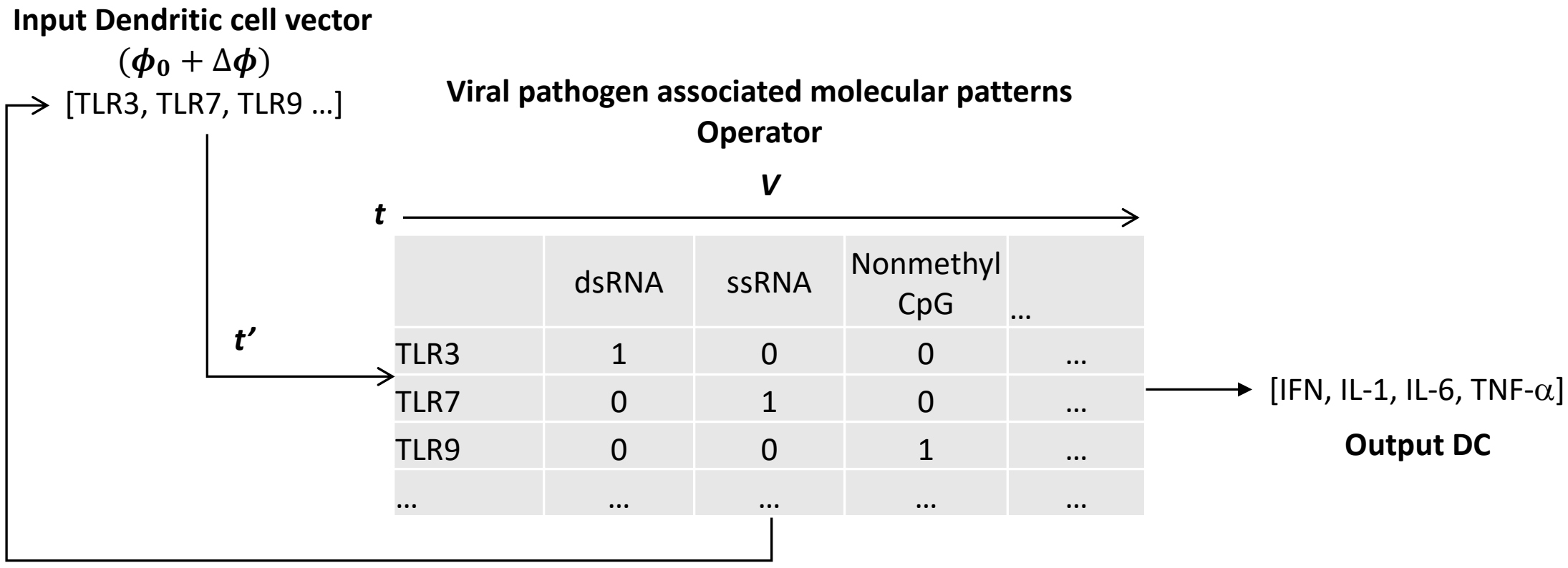

| | dsRNA | ssRNA | Nonmethyl CpG | … |
|---|---|---|---|---|
| TLR3 | 1 | 0 | 0 | … |
| TLR7 | 0 | 1 | 0 | … |
| TLR9 | 0 | 0 | 1 | … |
| … | … | … | … | … |

1/0 – Present/absent TLR affinity for PAMP. The term for DC is multiplied into this number, if V is persistently elevated the DC continues unabated, if V declines DC does as well

$$DC \approx (\phi_0 + \Delta\phi)e^{ct'}V$$

**B.** Differential impact of cytokines interactions with cytokine receptors with + and – growth effects on single immune effector cells.

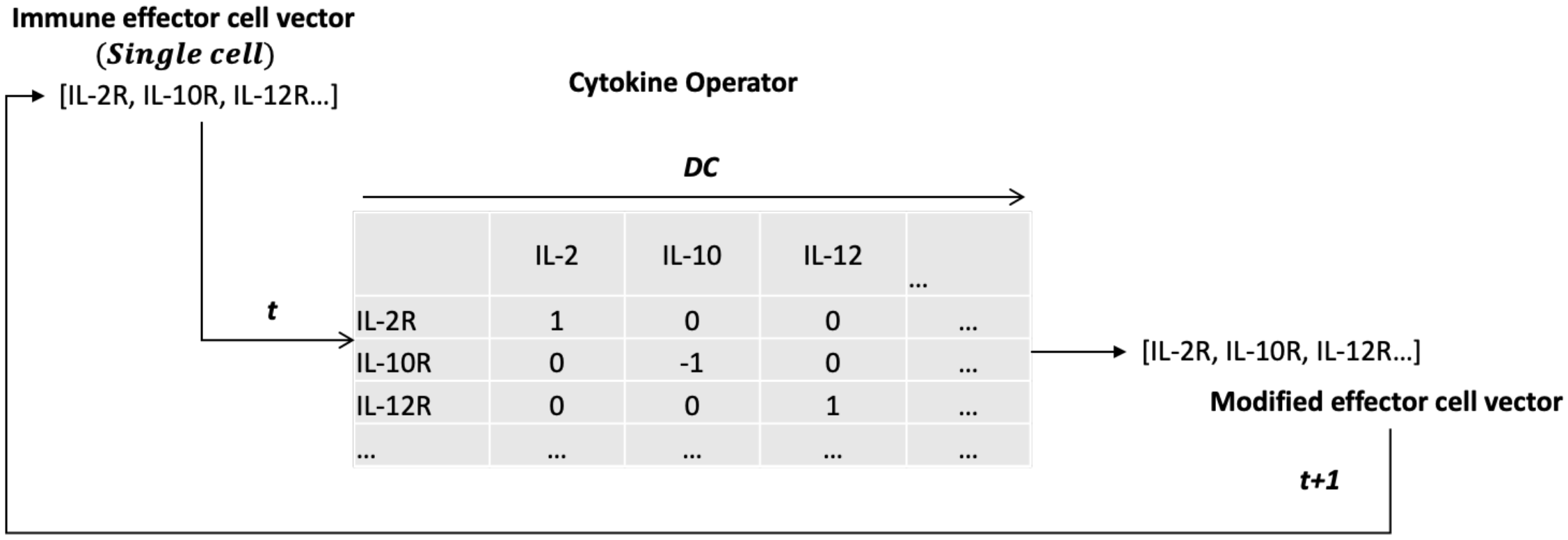

| | IL-2 | IL-10 | IL-12 | … |
|---|---|---|---|---|
| IL-2R | 1 | 0 | 0 | … |
| IL-10R | 0 | -1 | 0 | … |
| IL-12R | 0 | 0 | 1 | … |
| … | … | … | … | … |

1/0 – Present/absent cytokine affinity for receptor. The resulting cytokine-receptor interaction is a multiplier for the growth equation and goes into the growth rate exponent as well, modifying proliferation constant as well as growth. Depending on the relative concentration of cytokines, cell growth may be promoted or suppressed.

**C.** Differential impact of cytokines on immune effector cell subsets with + and – growth effects.

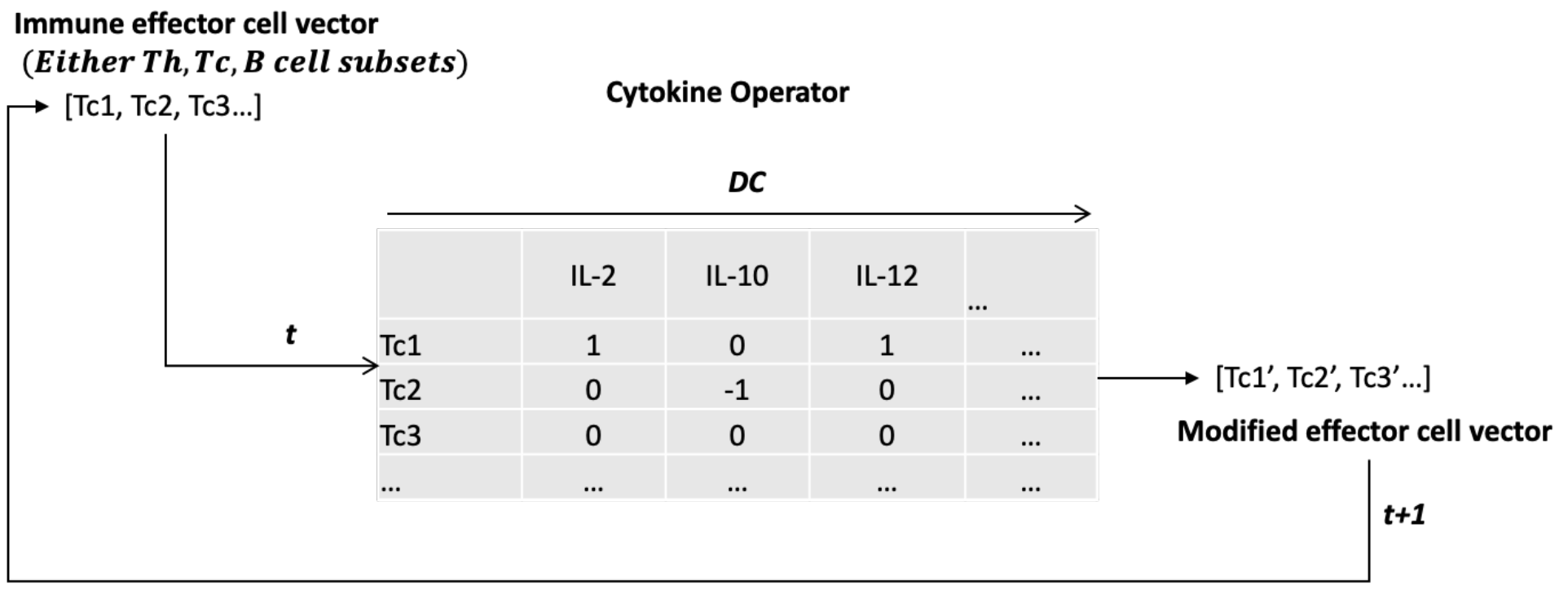

1/0 – Present/absent cytokine effect on T cell proliferation, each T cell has multiple receptors, which after interaction may either cause growth (+) or suppression (-).

*In this case*

$$Tc1<Tc1';\quad Tc2>Tc2';\quad Tc3=Tc3'$$

**D.** Differential impact of viral antigens binding affinity to HLA class I, II and BCR on immune effector cell clonal growth.

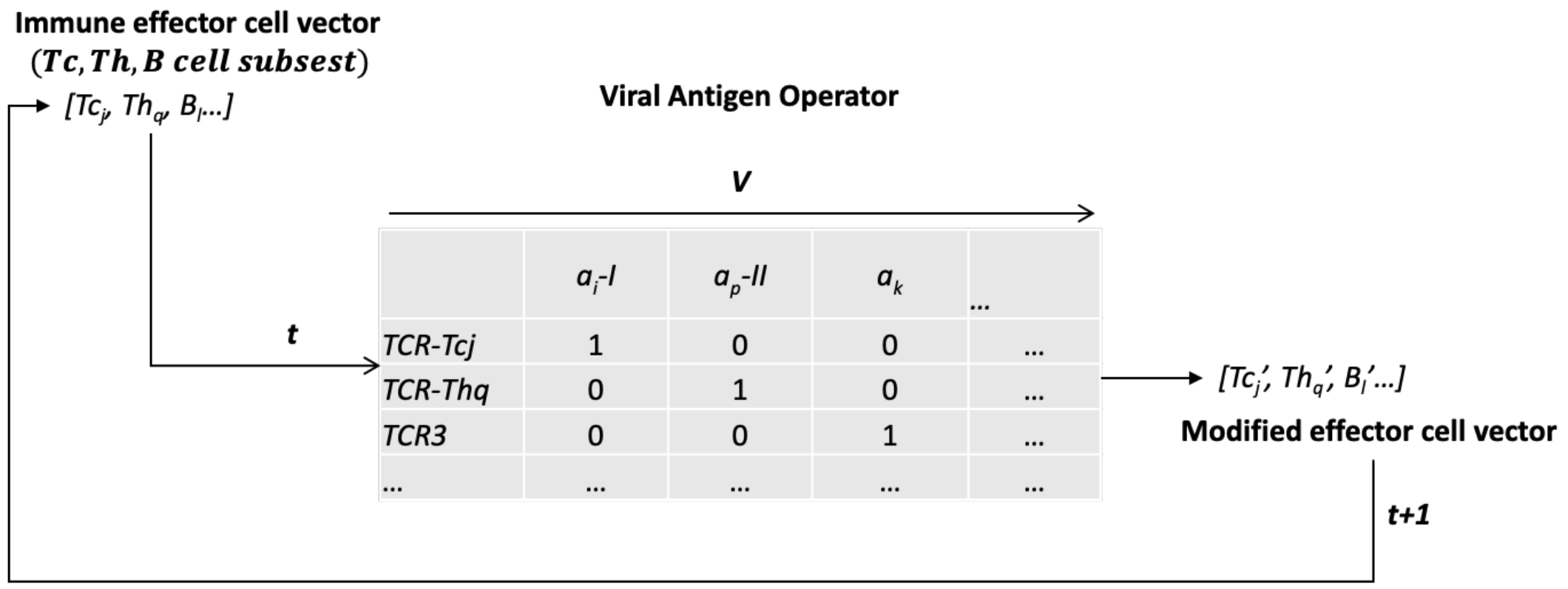

1/0 – Present/absent affinity of antigen – HLA complex or antigen for TCR or BCR respectively. Each immune effector will grow in proportion to the affinity and following the logistic equation of growth

**Figure 5.** Schematic showing the two scenarios, one with overlapping genetic traits vs. distinct genetic traits, with differential susceptibility to COVID-19.

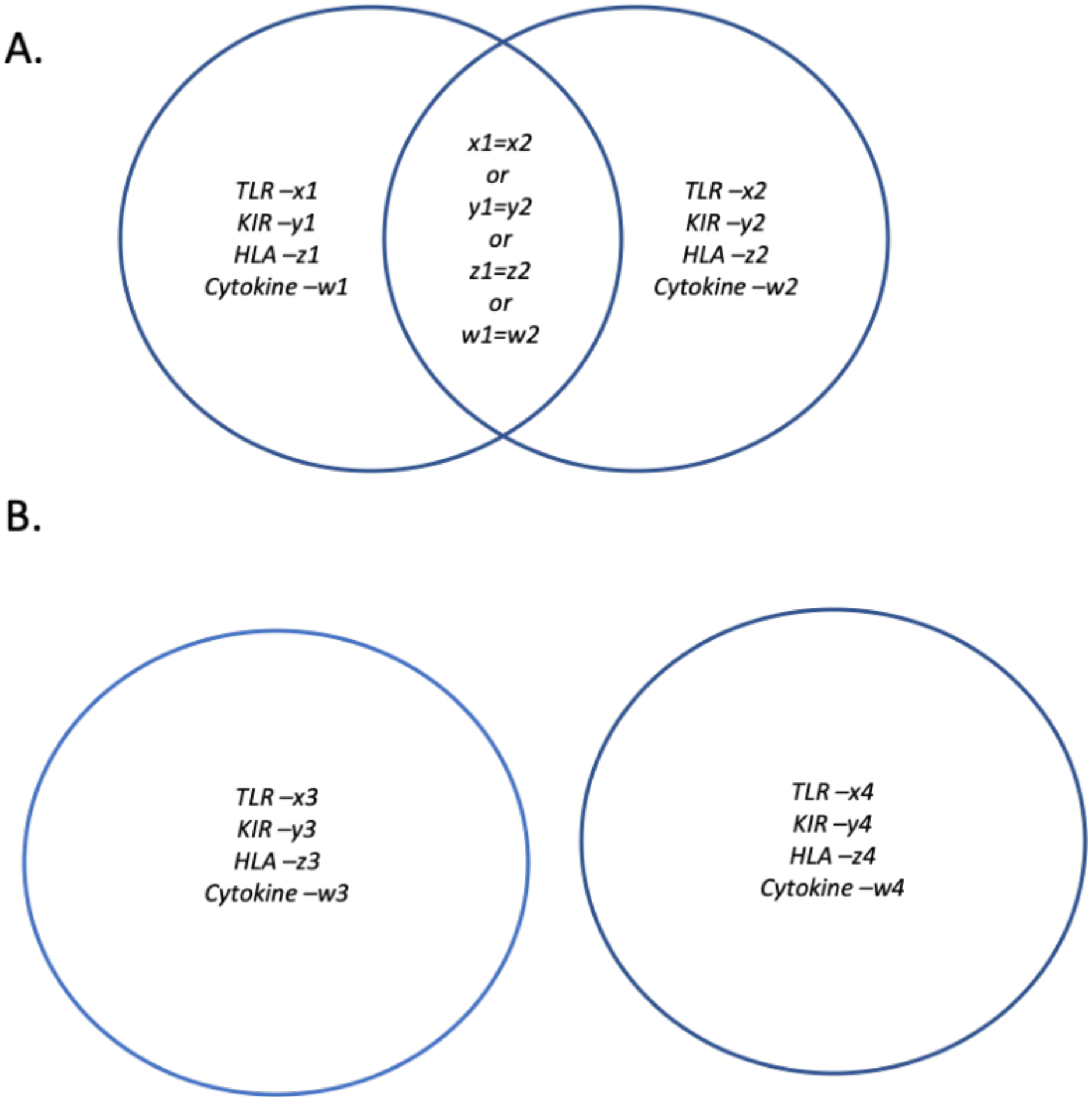

**Supplementary Table 1.** List of polymorphisms in immune regulatory genes.

| Cytokine | Polymorphism | Associated syndrome |
|---|---|---|
| **Interferon alpha** | IFNL3- rs8099917<br>Ss46915590<br>Rs117648444[1]<br>Rs368234815<br>IL28B<br>Ss469415590<br>Rs12979860 | HCV- response to IFN and ribavirin treatment[4] |
| | Rs1332190 (IFNA1)<br>RS 303218[2] | Chronic hepatitis B[5] |
| | Rs2843710 (IFN AR1)[3] | Enterovirus 71 |
| | Rs2243594 A/G<br>Rs1012335 G/C<br>Rs2257167 G/C<br>Rs 2843710 C/G<br>Rs2850015 C/T<br>+6993 C/T | Plasmodium falciparum malaria[6] |
| | Rs1332190 (IFNA1)<br>Rs9298814 (IFNA17) | Crimean-Congo Hemorrhagic Fever[7] |
| | IFNAR1_30127 (rs2254315, rs17875857)<br>IFNAR1_18339 (rs2257167, rs17875817) | HIV[8] |
| | JAK1<br>Rs11208534<br>Rs2780831<br>Rs310196<br>Rs669260 (DDX58) | Dengue[9] |
| | Rs368234815 | Increased risk of TRM after HCT[10] and viral reactivation after autologous HCT[11] |
| **Interferon beta** | Rs2304237<br>Rs2071430<br>Rs17000900 | SARS-CoV-1[12] |
| **Interferon gamma** | Rs12979860<br>Rs368234815 | Hepatitis C |
| | Rs2069705 T/G | Plasmodium falciparum malaria |
| | Rs2430561 A/T<br>Rs3138557 (CA)n<br>Rs2069718 T/C<br>Rs2068727 A/G<br>Rs2069728 G/A | |
| | Rs1859330 (OAS3) | Enterovirus 71[14,15] |
| | IFNGR1T-56-C | Non-tuberculous Mycobacteria[16] |
| | IFGN -1616, +3234<br>IFN -2109 A/G[13] | *Mycobacteria tuberculosis*[17] |
| | Rs2430561 | *Chlamydia trachomatis*[18]<br>*Mycobacteria tuberculosis*[19,20]<br>Hepatitis B[21]<br>Hepatitis C[22]<br>Cytomegalovirus[23,24]<br>BK virus[25]<br>Leprosy[26]<br>Chagas Disease[27]<br>Toxoplasmosis retinochoroiditis[28]<br>Dengue[29] |
| | Rs12369470<br>Rs 2406918 | BK virus[25] |
| | Rs3888188 (IFITM3)<br>Rs1861494<br>Rs2069718<br>Rs2430561 | *Mycobacteria tuberculosis*[30,31] |
| | OAS1-rs10774671, rs34137742 | West Nile Virus[29] |
| | Rs2430561 | Increased severity of GVHD after HCT[32] |
| **HLA** | HLA-DOA rs1044429<br>HLADOB rs2284191<br>Rs2856997<br>HLA-DMA rs1063478<br>HLA-DMB rs23544GG<br>Rs4273729 | Hepatitis C[33-35] |
| | HLA-B*46:01 | SARS-CoV-2, SARS CoV-1[36] |
| | HLA-B*15:18 | HIV[37-39] |
| | HLA-B*44:02<br>HLA-B*67:01<br>HLA-B*15:01G<br>HLA-DRB1*14:01<br>HLA- DRB1*13:02<br>HLA- DRB1*12:01<br>HLA-A*02:06 | |
| | HLA-A*25:01<br>HLA-C*01:02 | SARS-CoV-2[36] |
| | HLA-A*11<br>HLA-B*35<br>HLR-DRB1*10 | H1N1 Influenza[40] |
| | HLA-A*0203<br>HLA-A*24<br>HLA-A*33<br>HLA-DRB1*03<br>HLA-DRB1*09<br>HLA- DRB1*11 | Dengue[41] |
| | HLA-B*46<br>HLA-B*46-DRB1*09<br>HLA-B*51-DRB1*09 | Hantaan virus[42] |
| | Rs872956 (HLA-DPB1)<br>Rs9277535A (HLA-DPB1)<br>HLA-B*15<br>HLA-DRB1*11<br>HLA-DRB1*14<br>HLA-DQ rs9275572, rs2856718<br>HLA-DQB1&05:02<br>HLA-DQB1*05:03<br>HLA-DQB1*0601 | Hepatitis B[43-47] |
| | HLA-DRB1*030101<br>HLA-DRB1*130101<br>HLA- DRB1*040101<br>HLA-DRB1*040501<br>HLA- DRB1*7:01:01<br>HLA- DRB1*110101<br>HLA-DQ rs9275572, rs2856718<br>HLA-G rs1063320 | Hepatitis C[46,48,49] |
| | HLA-DQB1*0303 | *Helicobacter pylori*[50], Chikungunya[51] |
| | HLA-DQB1*0401 | *Helicobacter pylori*[50], Hepatitis B[47] |
| | HLA-DQA1*0103<br>HLA-DQ*0301 | *Helicobacter pylori*[50] |
| | HLA-E*0101/0101<br>HLA-G rs16375 | Cytomegolovirus[52,53] |
| **TLR** | TLR2 Arg753Gln<br>TLR2 rs5743709<br>TLR9 rs187084<br>TLR4 rs4986781 | *Mycobacteria tuberculosis*[54-57] |
| | TLR2 delta 22 | Plasmodium falciparum[58] |
| | TLR4 asp299gly | Influenza[59], HIV[60], cytomegalovirus[61] |
| | TLR2 arg677trp, | cytomegalovirus[61] |
| | TLR arg753gln | Influenza[59], Echinococcus[62],<br>*Mycobacteria tuberculosis*[63]<br>Cytomegalovirus[59] |
| | TLR3 leu412phe | Influenza[59], Cytomegalovirus[64],<br>Japanese Encephalitis Virus[65] |
| | TLR4 | Respiratory Syncytial Virus[66] |
| | TLR4 rs4986790<br>TLR4 rs4986791<br>TLR4 rs2149356 | Hepatitis C[67,68] |
| | TLR4 399 C | Hepatitis E[69] |
| | TLR4 rs11536889 | Hepatitis A[70] |
| | TLR-7 (rs3853839)<br>TLR-7 IVS1+1817G/T<br>TLR-7 c.4-141A/G<br>‘ | Chikungunya[71]<br>Crimean Congo Hemorrhagic fever[72] |
| | TLR-8 (rs 3764879) | Chikungunya[71] |
| | TLR-9 (rs352139) | *Plasmodium falciparum*[73] |
| | TLR-1 rs5743551 G/G<br>TLR-8 rs3764879 | Increased TRM after HCT[74] |
| | TLR-1 rs5742611 | Increased grade II-IV aGVHD after HCT[75] |
| | TLR-4 +3725G/G | Decreased TRM after HCT[76] |
| **IL1** | TIRAP 180L<br><br>IL-1B +3954 C/T<br><br>IL-1RN*1/*1<br>IL-1RN*2<br><br>IL-1B +3954 C/T | Recurrent pneumococcal LRTI[77]<br><br>sepsis (protective)[78]<br><br>Chikungunya[79]<br><br><br>Increased aGVHD after HCT[80] |
| **IL6** | Rs1800795<br><br>Rs1800796<br><br>Rs1800795<br>Rs1800797 | Septic shock from pneumonia[81]<br><br>*Mycobacteria tuberculosis*[82]<br><br>Increased rate of aGVHD after HCT[83,84] |
| **KIR** | KIR 2DL3<br>KIR 2DL5<br>KIR 2DL2<br>KIR 2DS1<br>KIR 3DS1<br><br>Haplotype B/x<br>KIR 2DL2 and 2DS2 | HIV [85,86]<br><br><br><br><br><br>Increased cGVHD after HCT[87]<br>Increased aGVHD after HCT[88] |
| **TNF alpha** | TNF857 C/C<br>TNF 238 A/A<br><br>TNF -308A<br><br><br><br><br><br><br>TNF 863C<br><br>rs1800629<br>rs1799724<br><br>857 C/T<br><br>RS361525(A)<br><br>Rs1800629 | *Mycobacteria tuberculosis*[13]<br><br><br>Toxoplasmosis retinochoroiditis[28]<br>Septic shock from pneumonia[81]<br>Schistosomiasis[89]<br>Hepatitis B[90]<br>Post-operative sepsis[91]<br>Japanese Encephalitis Virus[92]<br><br>Japanese Encephalitis Virus[92]<br><br>Hepatitis C[93,94]<br><br><br>Hepatitis B[90]<br><br>Increased aGVHD risk after HCT[95]<br><br>Sepsis risk[96] |
| **NOD** | Rs2284358<br>Rs2970500 | Cytomegalovirus[97] |
| | Rs10267377 | Toxoplasmic retinochoroiditis[98] |
| | LRRKrs1873614<br>Rs1873613GG<br>RIPK rs40457, rs42490 | *Mycobacterium leprae*[99] |
| | NOD2/Card2 variants | Increased TRM after HCT[100] |

Abbreviations: aGVHD=acute graft versus host disease, cGVHD=chronic graft versus host disease, TRM=treatment related mortality, HCT=hematopoietic stem cell transplant, LRTI= lower respiratory tract infection, HIV = human immunodeficiency virus. SARS-CoV-2 = severe acute respiratory syndrome coronavirus 2.

**Supplementary Table 1 References:**

## References.